\documentclass[aps,prb,twocolumn,showpacs]{revtex4}

\begin{document}
\title{Optical activity of noncentrosymmetric metals}

\author{V. P. Mineev$^1$ and Yu Yoshioka$^2$}

\affiliation{$^{1}$ Commissariat \`a l'Energie Atomique,
INAC/SPSMS, 38054 Grenoble, France\\ $^{2}$ Osaka University, Graduate School of Engineering Science, Toyonaka, Osaka 560-8531, Japan}
\date{\today}

\begin{abstract}
We describe the phenomenon of optical activity of noncentrosymmetric metals 
in their normal and  superconducting states. 
The found conductivity tensor contains the linear in wave vector off diagonal part 
responsible for the natural optical activity. 
Its value is expressed through 
the ratio of light frequency to the band splitting due to the spin-orbit interaction.
The Kerr rotation of polarization of light reflected from the metal surface is calculated.
\end{abstract}


\maketitle

\section{Introduction}
The metals without inversion symmetry have recently become a subject of considerable interest 
mostly due to the discovery of superconductivity in CePt$_3$Si.
\cite{Bauer04}
Now the list of noncentrosymmetric superconductors has grown to
include UIr \cite{Akazawa04}, CeRhSi$_3$ 
\cite{Kimura05}, CeIrSi$_3$ \cite{Sugitani06},
Y$_2$C$_3$ \cite{Amano04},
Li$_2$(Pd$_{1-x}$,Pt$_x$)$_3$B \cite{LiPt-PdB},
KOs$_2$O$_6$ \cite{KOsO}, and other compounds. 
The spin-orbit coupling of electrons in noncentrosymmetric crystal lifts the spin  degeneracy of the electron energy band causing a noticeable band splitting. The Fermi surface splitting can be observed by the de Haas-van Alphen effect discussed theoretically in the paper Ref. \onlinecite{MinSam05}.  The band splitting reveals itself in the large residual value of the spin susceptibility of noncentrosymmetric superconductors at zero temperature.\cite{Sam07} It also makes possible the existence of nonuniform superconducting states that can be traced to the Lifshitz invariants in the free energy.\cite{MinSam08}
 Another significant
manifestation of the band splitting is the natural optical activity.  

The natural optical activity or natural gyrotropy is well known phenomenon typical for the bodies having no centre of symmetry \cite{LL}. The optical properties of a naturally active 
body resemble those of the magneto-active media having no time reversal symmetry. It exhibits double circular refraction, the Faraday and the Kerr effects. In this case, the tensor of dielectric permeability  has  linear terms in the expansion in powers of wave vector
\begin{equation}
\varepsilon_{ij}(\omega,{\bf q})=\varepsilon_{ij}(\omega,0)+i\gamma_{ijl}(\omega)q_l,
\label{e1}
\end{equation}
where $\gamma_{ikl}(\omega)$ is an antisymmetric third rank tensor called the tensor of gyrotropy. 

The spacial dispersion term in permeability has been derived  by Arfi and Gor'kov \cite{Arfi} 
 in frame of  model where "a conductor lacking a center of inversion is simulated by an ordered arrangement of impurities whose scattering potential is asymmetric" (see also \cite{Levitov}). 
We shall be interested in gyrotropy properties of a clean noncentrosymmetric metal. 
In  metals, it is more natural to describe  optical properties in terms of spacial dispersion of conductivity tensor having the following form:
\begin{equation}
\sigma_{ij}(\omega,{\bf q})=\sigma_{ij}(\omega,0)-i\lambda_{ijl}(\omega)q_l.
\label{e2}
\end{equation}
Here, the gyrotropy tensor $\lambda_{ijl}(\omega)$ is an odd function of frequency.

The gyrotropic tensor has the most simple form in the metals with cubic symmetry.
In this case, the Drude part of the conductivity tensor is isotropic 
$\sigma_{ij}(\omega,0)=\sigma(\omega)\delta_{ij}$ and the gyrotropic conductivity tensor $\lambda_{ikl}
(\omega)=\lambda(\omega) e_{ikl}$ is determined by the single complex function  
 $\lambda(\omega)=\lambda'(\omega)+i\lambda''(\omega)$ 
 such that a 
normal state density of current  is 
\begin{equation}
{\bf j}=\frac{\epsilon}{4\pi}\frac{\partial {\bf E}}{\partial t}+{\sigma}{\bf E}+\lambda\text{rot}{\bf E}.
\label{cur}
\end{equation}
In time representation $\lambda$ is an operator being an odd function of operation of time derivative $\partial/\partial t$. 

In the superconducting state besides $\lambda\text{rot}{\bf E}$ the gyrotropic part of current density contains also an additional term proportional to the magnetic induction ${\bf B}$
\begin{equation}
{\bf j}_g=\lambda\text{rot}{\bf E}+\nu{\bf B}.
\label{g}
\end{equation}
Here $\nu=\nu(T)$ is a constant coefficient being equal to zero in the normal state.\cite{Levitov2,Yip2}

In this paper  
we present the derivation of 
 current response to the electro-magnetic field with finite frequency and wave vector valid for the normal and the superconducting state of noncentrosymmetric metals with cubic symmetry. 
We find that the gyrotropy conductivity $\lambda$ is directly proportional to the ratio of the light frequency to the energy of the band splitting.   
Then making use the Maxwell equations and corresponding boundary conditions at the surface of noncentrosymmetric  metal derived in the Section V we calculate
the Kerr rotation of polarization of light reflected from the surface of metal.


\section{current response to electro-magnetic field}

 The current response  of a clean metal to the electromagnetic field at finite 
${\bf q}$ and $\omega$ can be written following the textbook procedure \cite{FizKin}.
 In application to our situation one has to remember that due to spin-orbital coupling 
 determined by the dot product of the Pauli matrix vector $\mbox{\boldmath$\sigma$}$ and  pseudovector $\mbox{\boldmath$\gamma$}({\bf k})$, which is  odd in respect to momentum
 $\mbox{\boldmath$\gamma$}(-{\bf k})=-$\mbox{\boldmath$\gamma$}({\bf k})$ $  and specific for each noncentrosymmetric crystal structure \cite{Sam07,MinSig}, all the values such as single electron energy 
\begin{equation}
\xi_{\alpha\beta}({\bf k})=(\varepsilon({\bf k})-\mu)\delta_{\alpha\beta}+
\mbox{\boldmath$\gamma$}({\bf k})\mbox{\boldmath$\sigma$}_{\alpha\beta},
\end{equation}
velocity
${\bf v}_{\alpha\beta}({\bf k})=\partial\xi_{\alpha\beta}({\bf k})/\partial{\bf k}$, the inverse effective mass
$(m^{-1}_{ij})_{\alpha\beta}=\partial^2\xi_{\alpha\beta}({\bf k})/\partial k_i\partial k_j$, the Green functions 
$G_{\alpha\beta}(\tau_1,{\bf k};\tau_2,{\bf k}')=-\langle T_\tau a_{{\bf k}\alpha}(\tau_1)a^\dagger_{{\bf k}'\beta}(\tau_2)\rangle$ and $F_{\alpha\beta}(\tau_1,{\bf k};\tau_2,{\bf k}')=\langle T_\tau a_{{\bf k}\alpha}(\tau_1)a_{-{\bf k}'\beta}(\tau_2)\rangle$ are matrices in the spin space. 
Taking this in mind, we obtain
\begin{eqnarray}
\label{basic}
 j_i(\omega_n,{\bf q})= -\frac{e^2}{c}Tr\left [\hat m^{-1}_{ij}\hat n_e\right. +~~~~~~~~~~~~~\nonumber\\
\int  \frac{d^3k}{(2\pi\hbar)^3}\text{T}\sum_{m=-\infty}^{\infty} 
\{  \hat v_i({\bf k})\hat G^{(0)}(K_+)\hat v_j({\bf k})\hat G^{(0)}(K_-)\nonumber\\
\left.
+ \hat v_i({\bf k})\hat F^{(0)}(K_+)\hat v_j^t(-{\bf k})\hat F^{+(0)}(K_-)\}
\right ]{A_j}(\omega_n,{\bf q}).
\end{eqnarray}
The transposed matrix of velocity is determined as $\hat{\bf v}^t(-{\bf k})=\partial\hat \xi^t(-{\bf k})/\partial{\bf k}$.  The arguments of the zero field Green functions are denoted as $
K_{\pm}=\left(\Omega_m\pm{\omega_n}/{2}, {\bf k}\pm{\bf q}/{2}\right)$. The Matsubara frequencies 
take the values $\Omega_m=\pi (2m+1-n)T$ and $\omega_n=2\pi nT $. 


One can pass from the spin to the band representation, where the one-particle Hamiltonian
\begin{equation}
H_0=\sum_{\bf k}\xi_{\alpha\beta}({\bf k})a^{\dagger}_{{\bf k}\alpha}a_{{\bf k}\beta}=
\sum_{{\bf k},\lambda=\pm}\xi_{\lambda}({\bf k})c^{\dagger}_{{\bf k}\lambda}c_{{\bf k}\lambda}
\end{equation}
has diagonal form. Here, the band energies are 
\begin{equation}
\xi_{\lambda}({\bf k})=\varepsilon({\bf k})-\mu+\lambda
|\mbox{\boldmath$\gamma$}({\bf k})|,
\end{equation}
such that two Fermi surfaces are determined by equations $\xi_{\lambda}({\bf k})=0 $.
The difference of the band energies $2|\mbox{\boldmath$\gamma$}({\bf k}_F)|$ characterizes   the intensity of the spin-orbital coupling. The Fermi momentum with $\mbox{\boldmath$\gamma$}=0$ is determined by the equation $\varepsilon({\bf k}_F)=\varepsilon_F$.

The diagonalization is made by the following transformation
\begin{equation}
\label{band transform}
    a_{{\bf k}\alpha}=\sum_{\lambda=\pm}u_{\alpha\lambda}({\bf k})c_{{\bf k}\lambda},
\end{equation}
with the coefficients
$$
    \begin{array}{l}
    \displaystyle u_{\uparrow\lambda}{(\bf k})=
    \sqrt{\frac{|\mbox{\boldmath$\gamma$}|+\lambda\gamma_z}{2|\mbox{\boldmath$\gamma$}|}},\\
    \displaystyle u_{\downarrow\lambda}({\bf k})=\lambda
    \frac{\gamma_x+i\gamma_y}{\sqrt{2|\mbox{\boldmath$\gamma$}|(|\mbox{\boldmath$\gamma$}|+\lambda\gamma_z)}},  
\end{array}
$$
forming a unitary matrix $\hat u({\bf k})$.

The zero field Green functions in the band representation are diagonal and have the following form:
\cite{Sam07}
\begin{eqnarray}
G^{(0)}_{\lambda\lambda'}(\omega_n,{\bf k})=\delta_{\lambda\lambda'}G_{\lambda}(\omega_n,{\bf k}),\nonumber\\
F^{(0)}_{\lambda\lambda'}(\omega_n,{\bf k})=\delta_{\lambda\lambda'}F_{\lambda}(\omega_n,{\bf k}),
\end{eqnarray} 
where
\begin{eqnarray}
G_{\lambda}(\omega_n,{\bf k})=-\frac{i\hbar\omega_n+\xi_\lambda}{\hbar^2\omega_n^2+\xi_\lambda^2+
|\tilde\Delta_{\lambda}({\bf k})|^2},\nonumber\\
F_{\lambda}(\omega_n,{\bf k})=\frac{t_\lambda({\bf k})\tilde\Delta_{\lambda}({\bf k})}{\hbar^2\omega_n^2+\xi_\lambda^2+
|\tilde\Delta_{\lambda}({\bf k})|^2},
\end{eqnarray} 
and
$$
    t_{\lambda}({\bf k})=-\lambda
    \frac{\gamma_x({\bf k})-i\gamma_y({\bf k})}{\sqrt{\gamma_x^2({\bf k})+\gamma_y^2({\bf k})}}.
$$
The functions $\tilde\Delta_{\lambda}({\bf k})$ are the gaps in the $\lambda$-band quasiparticle spectrum in superconducting state.
In the simplest model with BCS pairing interaction $v_g({\bf k},{\bf k}')= -V_g$, the gap functions are the same in both bands: $\tilde\Delta_{+}({\bf k})=\tilde\Delta_{-}({\bf k})=\Delta$ and we deal with pure singlet pairing \cite{SamMin08}.

 Transforming the Green functions using eqn. (\ref{band transform}) into the band representation
 \cite{FaVa}, we obtain
 \begin{eqnarray}
 \label{trace}
Tr \{  \hat v_i({\bf k})\hat G^{(0)}(K_+)\hat v_j({\bf k})\hat G^{(0)}(K_-)\nonumber\\
+ \hat v_i({\bf k})\hat F^{(0)}(K_+)\hat v_j^t(-{\bf k})\hat F^{+(0)}(K_-)\}=\nonumber\\
v_{++,i}({\bf k})G_+v_{++,j}({\bf k})G_++v_{++,i}({\bf k})F_+v_{++,j}(-{\bf k})F^\dagger_++\nonumber\\
v_{--,i}({\bf k})G_-v_{--,j}({\bf k})G_-+v_{--,i}({\bf k})F_-v_{-+,j}(-{\bf k})F^\dagger_++\nonumber\\
v_{+-,i}({\bf k})G_-v_{-+,j}({\bf k})G_++v_{+-,i}({\bf k})F_-v_{+-,j}(-{\bf k})F^\dagger_++\nonumber\\
v_{-+,i}({\bf k})G_+v_{+-,j}({\bf k})G_-+v_{-+,i}({\bf k})F_+v_{-+,j}(-{\bf k})F^\dagger_-~~
 \end{eqnarray}
as the trace of the matrices in eqn. (\ref{basic}).
For the brevity, we omit here the arguments of the Green functions.
They are the same as in the upper two lines.
The matrix velocity in the band representation is 
 \begin{eqnarray}
 {\bf v}_{\lambda\lambda'}(\pm{\bf k})=u^{\dagger}_{\lambda\alpha}(\pm{\bf k}){\bf v}_{\alpha\beta}(\pm{\bf k})u_{\beta\lambda'}(\pm{\bf k})=\nonumber\\
 \frac{\partial \xi_0(\pm{\bf k})}{\partial{\bf k}}\delta_{\lambda\lambda' }+
\frac{\partial\gamma_l(\pm{\bf k})}{\partial{\bf k}}\tau_{l,\lambda\lambda'}(\pm{\bf k}),
\label{speed}
 \end{eqnarray}
 where 
$\mbox{\boldmath$\hat \tau$}({\bf k})
=\hat u^{\dagger}({\bf k})\mbox{\boldmath$\hat\sigma$}\hat u({\bf k})$ are hermitian matrices. We have neglected \cite{neglect}  the difference between $\hat u({\bf k})$ and $\hat u({\bf k}\pm{\bf q}/2)$.

The explicit expressions for the $\mbox{\boldmath$\hat \tau$}({\bf k})$ matrices  are
\begin{eqnarray}
\label{m_i}
    &&\hat \tau_x=
    \left(\begin{array}{cc}
      \hat\gamma_x & -
      \frac{\gamma_x\hat\gamma_z+i\gamma_y}{\gamma_\perp} \\
      -
      \frac{\gamma_x\hat\gamma_z-i\gamma_y}{\gamma_\perp} & -\hat\gamma_x \\
    \end{array}\right),\nonumber\\
    &&\hat \tau_y=
    \left(\begin{array}{cc}
      \hat\gamma_y & -
      \frac{\gamma_y\hat\gamma_z-i\gamma_x}{\gamma_\perp} \\
      -
      \frac{\gamma_y\hat\gamma_z+i\gamma_x}{\gamma_\perp} & -\hat\gamma_y \\
    \end{array}\right),\quad\\
    &&\hat \tau_z=
    \left(\begin{array}{cc}
      \hat\gamma_z & \frac{\gamma_\perp}{\gamma} \\
      \frac{\gamma_\perp}{\gamma} & -\hat\gamma_z \\
    \end{array}\right)\nonumber,
\end{eqnarray}
where $\hat{\mbox{\boldmath$\gamma$}}=\mbox{\boldmath$\gamma$}/|\mbox{\boldmath$\gamma$}|$,
$\gamma_\perp=\sqrt{\gamma_x^2+\gamma_y^2}$.  All the diagonal elements of these matrices are odd functions of the momentum components. Correspondigly, the products of the diagonal components of velocity matrices (\ref{speed}) are even functions of momentum. Hence, these terms in eqn. (\ref{trace}) can 
produce only the terms proportional to  even powers of the product ${\bf k}{\bf q}$ being not responsible for the gyrotropy properties.

The off diagonal elements of $\mbox{\boldmath$\hat \tau$}({\bf k})$ have the mixed parity. So, the  dispersive  terms proportional to the odd powers of the  product 
${\bf k}{\bf q}$ can arise only from the part of eqn. (\ref{trace}) containing the off diagonal elements of  
$\mbox{\boldmath$\hat \tau$}({\bf k})$ matrices.
Hence, for the calculation of gyrotropy of conductivity, only the last two lines in eqn. (\ref{trace}) consisting of interband terms are important. They are equal to
\begin{eqnarray}
 \label{trace'}
\frac{\partial\gamma_l}{\partial k_i}
\frac{\partial\gamma_m}{\partial k_j}\{\tau_{l,+-}\tau^{*}_{m,+-}
[G_-G_+-F_-F^\dagger_+]+\nonumber\\
\tau_{l,-+}\tau^{*}_{m,-+}
[G_+G_--F_+F^\dagger_-]\}.
 \end{eqnarray}
Using the identity 
$$
\tau_{l,+-}\tau^{*}_{m,+-}=\tau^*_{l,-+}\tau_{m,-+}=\delta_{lm}-\hat\gamma_l\hat\gamma_m+ie_{lmn}\hat\gamma_n,
$$
where $\hat\gamma_l=\gamma_l/|\mbox{\boldmath$\gamma$}|$,
one can rewrite the above expression as
\begin{eqnarray}
 \label{trace1}
\frac{\partial\gamma_l}{\partial k_i}
\frac{\partial\gamma_m}{\partial k_j}\{(\delta_{lm}-\hat\gamma_l\hat\gamma_m)
[G_-G_++G_+G_--F_-F^\dagger_+~~~~~~~\\
-F_+F^\dagger_-]+
ie_{lmn}\hat\gamma_n
[G_-G_+-G_+G_--F_-F^\dagger_++F_+F^\dagger_-]\}.\nonumber
 \end{eqnarray}
Starting this point we need the explicit form of spin-orbit coupling vector $\mbox{\boldmath$\gamma$}({\bf k})$. Its momentum dependence  is determined by the
crystal symmetry.\cite{MinSig, Sam07} For the cubic group $G=O$,
which describes the point symmetry of
Li$_2$(Pd$_{1-x}$,Pt$_x$)$_3$B, the simplest form compatible with
the symmetry requirements is
\begin{equation}
\label{gamma_O}
 \mbox{\boldmath$\gamma$}({\bf k})=\gamma_0{\bf k},
\end{equation}
where $\gamma_0$ is a constant. 
For the tetragonal group
$G={C}_{4v}$, which is relevant for CePt$_3$Si,
CeRhSi$_3$ and CeIrSi$_3$, the spin-orbit coupling is given by
\begin{equation}
\label{gamma C4v}
\mbox{\boldmath$\gamma$}({\bf k})=\gamma_\perp(k_y\hat x-k_x\hat y)
    +\gamma_\parallel k_xk_yk_z(k_x^2-k_y^2)\hat z.
\end{equation}

The gyrotropy current ${\bf j}_g$,  which is linear with respect to the wave vector ${\bf q}$, originates  from the last term in the eqn. (\ref{trace1}). 
One can show that for  the tetragonal crystal with the symmetry group $G={C}_{4v}$, for the electric field lying in the basal plane
 the linear  in the component of wave vector ${\bf q}$ part of conductivity is absent.  
In that follows we continue calculation for the metal with cubic symmetry where 
$\hat\gamma=\hat{k}~{sign}\gamma_0$.  
We put  $\hat\gamma=\hat{k} $ taking
$\gamma_0$ as a positive constant. Thus, we obtain for gyrotropy current \cite{total}
\begin{equation}
 \label{trace2}
j_{gi}(\omega_n,{\bf q})=ie_{ijl}\frac{e^2\gamma_0^2}{c}I_l{A_j}(\omega_n,{\bf q}),
\end{equation}
\begin{eqnarray}
I_l=\int  \frac{d^3k}{(2\pi\hbar)^3}\hat k_l~~~~~~~~~~~~~~~~~~~~~~~~~~~~~~~\nonumber\\
\times\text{T}\sum_{m=-\infty}^{\infty}
[G_+(K_+)G_-(K_-)-F_+(K_+)F^\dagger_-(K_-)-\nonumber\\ 
-G_-(K_+)G_+(K_-)+F_-(K_+)F^\dagger_+(K_-)].
\label{integ}
 \end{eqnarray}

Let us find first the gyrotropy conductivity in the normal state. 

\section{gyrotropy conductivity in the normal state}

Substituting the Green function in the eqn. (\ref{integ}) and 
performing summation over the Matsubara frequencies, we obtain
\begin{eqnarray}
I_l=\int  \frac{d^3k}{(2\pi\hbar)^3}\hat k_l
\left\{\frac{f(\xi_-({\bf k}_-))-f(\xi_+({\bf k}_+))}{i \hbar\omega_n+\xi_-({\bf k}_-)-\xi_+({\bf k}_+)}-
\right.\nonumber\\
-\left.\frac{f(\xi_+({\bf k}_-))-f(\xi_-({\bf k}_+))}{i\hbar\omega_n+\xi_+({\bf k}_-)-\xi_-({\bf k}_+)}\right\}. 
 \end{eqnarray}
Here $f(\xi_\pm( {\bf k}_{\pm}))$ is the Fermi distribution function and $ {\bf k}_{\pm}={\bf k}\pm {\bf q}/2$.
By changing  the sign of momentum  ${\bf k}\to-{\bf k} $ in the first term under integral and making use that $\xi_\lambda({\bf k})$ is even function of ${\bf k}$, we come to
\begin{equation}
I_l=2
\int  \frac{d^3k}{(2\pi\hbar)^3}\hat k_l
\frac{[\xi_+({\bf k}_+)-\xi_-({\bf k}_-)][f(\xi_+({\bf k}_+))
-f(\xi_-({\bf k}_-))]}
{(\xi_+({\bf k}_+)-\xi_-({\bf k}_-))^2-(i\hbar\omega_n)^2}.
\label{j}
 \end{equation}
Analytical continuation of this expression from the discrete set of Matsubara frequencies into entire 
half-plane $\omega>0$ is performed by the usual substitution $i\omega_n\to\omega+i/\tau$. 

We shall work at frequencies smaller when the band splitting $\hbar\omega<\gamma_0k_F$   far from the resonance region $\hbar\omega\approx\gamma_0k_F$   but still in the collisionless
limit $\omega\tau>1$ where one can decompose  the integrand in powers of $\omega^2$:
\begin{eqnarray}
I_{l}=2\int  \frac{d^3k}{(2\pi\hbar)^3}\hat k_l
[{f(\xi_+({\bf k}_+))
-f(\xi_-({\bf k}_-))}]\nonumber\\
\times\left\{\frac{1}{\xi_+({\bf k}_+)-\xi_-({\bf k}_-)}+ \frac{(\hbar\omega)^2}{(\xi_+({\bf k}_+)-\xi_-({\bf k}_-))^3 }\right\},
\label{J}
 \end{eqnarray} 
 The frequency independent term in eqn. (\ref{J}) corresponds to the current density $\nu{\bf B}$
introduced in eqn. (\ref{g}).
We are interested in linear in ${\bf q}$ part of density of current. 
Expanding the integrand 
up to the first order in $\frac{\partial\xi_\pm}{\partial{\bf k}}{\bf q}$ 
one can prove by direct calculation that this term vanishes.
Thus, in the normal state $\nu=0$ as it should be in gauge invariant theory (see Section V and  \cite{Levitov2}). 
 The frequency dependent term determines the current 
\begin{equation}
\label{I}
j^g_i(\omega,{\bf q})=ie_{ijl}\frac{e^2\gamma_0^2}{c}
\hbar q_m(\hbar\omega)^2 I_{lm}{A_j}(\omega,{\bf q}),~
\end{equation}
where
\begin{eqnarray}
\label{int_n}
I_{lm}=\int  \frac{d^3k}{(2\pi\hbar)^3}
\hat k_l\left [-3\frac{f(\xi_+)-f(\xi_-)}{(\xi_+-\xi_-)^{4}}\left ( \frac{\partial\xi_+}{\partial k_m}+\frac{\partial\xi_-}{\partial k_m} \right )
\right.+\nonumber\\
+\frac{1}{(\xi_+-\xi_-)^3}\left.\left (\frac{\partial f(\xi_+)}{\partial \xi_+}\frac{\partial\xi_+}{\partial{k_m}}+ 
\frac{\partial f(\xi_-)}{\partial \xi_-}\frac{\partial\xi_-}{\partial{k_m}}   \right )\right ].~~~~
 \end{eqnarray}

After substitution of the Fourier component of the vector potential by the Fourier component of an electric field
${\bf A}=c{\bf E}/i\omega$,  we obtain
\begin{equation}
j^g_i(\omega,{\bf q})=e_{ijl}e^2\hbar^3\gamma_0^2\omega q_mI_{lm}{E_j}(\omega,{\bf q}).
\label{J1}
 \end{equation} 
Performing integration  over momentum space for the spherical Fermi surfaces in the limit $\gamma_0 k_F\ll\varepsilon_F$, we obtain 
\begin{equation}
j^g_i(\omega,{\bf q})=e_{ijl}\frac{e^2\omega }{12\pi^2\gamma_0k_F}q_l{E_j}(\omega,{\bf q}).
\label{J'}
\end{equation}
The corresponding gyrotropy conductivity is
\begin{equation}
\lambda=i\frac{e^2\omega}{12\pi^2\gamma_0k_F}.
\label{lambda}
\end{equation}

\section{gyrotropy conductivity in the superconducting state}

To find the gyrotropy conductivity in the superconducting phase with the cubic symmetry, one needs to perform summation and integration in the  eqn. (\ref{integ}) using $G$ and $F$ in the superconducting phase.
The integral in eqn. (\ref{integ}) consists of two terms
\begin{equation}
\label{B1}
I_l=\int  \frac{d^3k}{(2\pi\hbar)^3}\hat k_l[J_{+-}({\bf k},\omega)-J_{-+}({\bf k},\omega)],
\end{equation}
where
\begin{equation}
\label{app1}
J_{+-}({\bf k},\omega)=T\sum_{m=-\infty}^\infty
   [G_+(K_+)G_-(K_-)-F_+(K_+)F_-^\dagger (K_-)],
\end{equation}
and
\begin{equation}
J_{-+}({\bf k},\omega)=T\sum_{m=-\infty}^\infty
   [G_-(K_+)G_+(K_-)-F_-(K_+)F_+^\dagger (K_-)].
\end{equation}

Transforming the summation into an equivalent contour integration \cite{FizKin},
eqn. (\ref{app1}) can be written as
\begin{eqnarray}
  \label{app2}
 J_{+-}({\bf k},\omega)= \frac{\hbar}{4\pi i} \oint d\omega'\tanh \frac{\hbar\omega '}{2T}\nonumber\\
 \times \left\{ [ G^R_+ (\omega',{\bf k}_+) -G^A_+ (\omega',{\bf k}_+) ]
              G^A_-(\omega' - \omega,{\bf k}_-)
         \right. \nonumber \\ 
         +[ G^R_- (\omega',{\bf k}_-) -G^A_- (\omega',{\bf k}_-) ]
              G^R_+(\omega' + \omega,{\bf k}_+)
         \nonumber \\
         -[ F^R_+ (\omega',{\bf k}_+) -F^A_+ (\omega',{\bf k}_+) ]
              F^A_-(\omega' - \omega,{\bf k}_-)
          \nonumber \\ \left.
         -[ F^R_- (\omega',{\bf k}_-) -F^A_- (\omega',{\bf k}_-) ]
              F^R_+(\omega' + \omega,{\bf k}_+)
         \right\}.
         \nonumber \\
\end{eqnarray}
Here,  the Green functions are
\begin{equation}
 G^{R,A}_\lambda(\omega,{\bf k})= \frac{u^2_\lambda({\bf k})}{\hbar\omega-\epsilon_\lambda({\bf k})
 \pm i\delta}
                +\frac{v^2_\lambda({\bf k})}{\hbar\omega+\epsilon_\lambda({\bf k})\pm i\delta},
 \end{equation}
 and
 \begin{equation}               
  F^{R,A}_\lambda(\omega,{\bf k})=\frac{t_\lambda({\bf k})\Delta}{2\epsilon_\lambda({\bf k})}
                \left[ \frac{1}{\hbar\omega+\epsilon_\lambda({\bf k})\pm i\delta}
                -\frac{1}{\hbar\omega-\epsilon_\lambda({\bf k})\pm i\delta}\right],
\end{equation}
where
\begin{equation}
 \left.u^2_\lambda({\bf k})\atop v^2_\lambda({\bf k})\right\} = \frac 12 \left( 1\pm \frac{\xi_\lambda({\bf k})}{\epsilon_\lambda({\bf k})} \right),
 \end{equation}
 \begin{equation}
 \epsilon^2_\lambda({\bf k}) = \xi^2_\lambda({\bf k}) + \Delta^2,
\end{equation}
and ${\bf k}_\pm = {\bf k}\pm {\bf q}/2$.
Taking into account
\begin{equation}
G^R_\lambda-G^A_\lambda=-2\pi i[u^2_\lambda\delta(\hbar\omega-\epsilon_\lambda)+v^2_\lambda\delta(\hbar\omega+\epsilon
_\lambda)],
\end{equation}
\begin{equation}
F^R_\lambda-F^A_\lambda=\frac{\pi it_\lambda\Delta}{\epsilon_\lambda}[\delta(\hbar\omega-\epsilon_\lambda)-\delta(\hbar\omega+\epsilon_\lambda)],
\end{equation}
after  integration with respect to $\omega'$, we can rewrite eqn. (\ref{app2}) as:
\begin{widetext}
\begin{eqnarray}
J_{+-}({\bf k},\omega)=-\frac{1}{2}
   \left[
    \left( \tanh \frac{\epsilon_+({\bf k_+})}{2T}
         - \tanh \frac{\epsilon_-({\bf k_-})}{2T}
    \right)
    \left( \frac{u^2_+({\bf k}_+)u^2_-({\bf k}_-)}
           {\epsilon_+({\bf k}_+)-\epsilon_-({\bf k}_-)-\hbar\omega}
         + \frac{v^2_+({\bf k}_+)v^2_-({\bf k}_-)}
           {\epsilon_+({\bf k}_+)-\epsilon_-({\bf k}_-)+\hbar\omega}
    \right) \right. \nonumber \\
    \left.
+   \left( \tanh \frac{\epsilon_+({\bf k_+})}{2T}
         + \tanh \frac{\epsilon_-({\bf k_-})}{2T}
    \right)
    \left( \frac{u^2_+({\bf k}_+)v^2_-({\bf k}_-)}
           {\epsilon_+({\bf k}_+)+\epsilon_-({\bf k}_-)-\hbar\omega}
         + \frac{v^2_+({\bf k}_+)u^2_-({\bf k}_-)}
           {\epsilon_+({\bf k}_+)+\epsilon_-({\bf k}_-)+\hbar\omega}
    \right) \right] \nonumber \\
-   \frac {1}{2}\frac{\Delta^2}{4\epsilon_+({\bf k}_+)\epsilon_-({\bf k}_-)}
    \left[-
    \left( \tanh \frac{\epsilon_+({\bf k_+})}{2T}
         - \tanh \frac{\epsilon_-({\bf k_-})}{2T}
    \right)
    \left( \frac{1}{\epsilon_+({\bf k}_+)-\epsilon_-({\bf k}_-)-\hbar\omega}
         + \frac{1}{\epsilon_+({\bf k}_+)-\epsilon_-({\bf k}_-)+\hbar\omega}
    \right) \right.  \nonumber \\
    \left.
+   \left( \tanh \frac{\epsilon_+({\bf k_+})}{2T}
         + \tanh \frac{\epsilon_-({\bf k_-})}{2T}
    \right)
    \left( \frac{1}{\epsilon_+({\bf k}_+)+\epsilon_-({\bf k}_-)-\hbar\omega}
         + \frac{1}{\epsilon_+({\bf k}_+)+\epsilon_-({\bf k}_-)+\hbar\omega}
    \right) \right].
\label{GGFF}
\end{eqnarray}
\end{widetext}
Here we have ignored  the shifts in the arguments of the phase factors: $t_\lambda({\bf k\pm{\bf q}/2})\approx t_\lambda({\bf k})$ leading to the  small corrections of the order of $\gamma_0k_F/\varepsilon_F$ to the main terms.

For the second term under integral in the eqn. (\ref{B1}) we have 
\begin{equation}
J_{-+}({\bf k},\omega)=J_{+-}(-{\bf k},-\omega).
\end{equation}
Hence,  the integral (\ref{B1}) can be rewritten as
\begin{equation}
I_l=\int  \frac{d^3k}{(2\pi\hbar)^3}\hat k_l[J_{+-}({\bf k},\omega)+J_{+-}({\bf k},-\omega)].
\end{equation}
It means that we should work with the doubled  even part of eqn. (\ref{GGFF}).
Expanding the integrand in powers of $\omega$ after long but straightforward calculations we come 
to the following formula
\begin{eqnarray}
I_{l}= 2\int  \frac{d^3k}{(2\pi\hbar)^3}\hat k_l[{n(\xi_+({\bf k}_+))
-n(\xi_-({\bf k}_-))}]~~~~~~~\nonumber\\
\times\left\{\frac{1}{\xi_+({\bf k}_+)-\xi_-({\bf k}_-)}+ \frac{(\hbar\omega)^2}{(\xi_+({\bf k}_+)-\xi_-({\bf k}_-))^3 }\right\}~~~~~\nonumber\\
-2\Delta^2\int  \frac{d^3k}{(2\pi\hbar)^3}\hat k_l\frac{(\hbar\omega)^2}{
(\xi_+({\bf k}_+)+\xi_-({\bf k}_-))(\xi_+({\bf k}_+)-\xi_-({\bf k}_-))^3 }\nonumber\\
\times\left (
\frac{\tanh\frac{\epsilon_+({\bf k}_+)}{2T}}
{\epsilon_+({\bf k}_+)}-\frac{\tanh\frac{\epsilon_-({\bf k}_-)}{2T}}
{\epsilon_-({\bf k}_-)}\right ),~~~~~~~~~~~~~~~~~~~~
\label{J_s}
 \end{eqnarray} 
where
\begin{equation}
n(\xi)=\frac{1}{2}\left(1-\frac{\xi}{\epsilon}\tanh\frac{\epsilon}{2T}   \right )
\end{equation}
is the  distribution function of electrons over energies. At $\Delta/\gamma_0k_F\ll 1$ the second integral is obviously much smaller than the first one.  So, we come to the expression
\begin{eqnarray}
I_{l}\cong 2\int  \frac{d^3k}{(2\pi\hbar)^3}\hat k_l
[{n(\xi_+({\bf k}_+))
-n(\xi_-({\bf k}_-))}]~~~~~~~\nonumber\\
\times\left\{\frac{1}{\xi_+({\bf k}_+)-\xi_-({\bf k}_-)}+ \frac{(\hbar\omega)^2}{(\xi_+({\bf k}_+)-\xi_-({\bf k}_-))^3 }\right\}
\label{J_{ss}}
\end{eqnarray} 
which has the same form as the corresponding formula for the normal state (\ref{J}).

Expanding the integrand 
up to the first order in $\frac{\partial\xi_\pm}{\partial{\bf k}}{\bf q}$ 
 we obtain for the current given by eqn. (\ref{trace2}):
 \begin{equation}
\label{I}
j^g_i(\omega,{\bf q})=ie_{ijl}\frac{e^2\gamma_0^2}{c}
\hbar q_m[I_{lm}^1+(\hbar\omega)^2 I_{lm}^3]{A_j}(\omega,{\bf q}),~
\end{equation}
\begin{eqnarray}
\label{int_s}
I_{lm}^1=\int  \frac{d^3k}{(2\pi\hbar)^3}
\hat k_l\left [-\frac{n(\xi_+)-n(\xi_-)}{(\xi_+-\xi_-)^2}\left ( \frac{\partial\xi_+}{\partial k_m}+\frac{\partial\xi_-}{\partial k_m} \right )
\right.\nonumber\\
+\frac{1}{\xi_+-\xi_-}\left.\left (\frac{\partial n(\xi_+)}{\partial \xi_+}\frac{\partial\xi_+}{\partial{k_m}}+ 
\frac{\partial n(\xi_-)}{\partial \xi_-}\frac{\partial\xi_-}{\partial{k_m}}   \right )\right ],~~~~
 \end{eqnarray}
\begin{eqnarray}
I_{lm}^3=\int  \frac{d^3k}{(2\pi\hbar)^3}
\hat k_l
\left [-3\frac{n(\xi_+)-n(\xi_-)}{(\xi_+-\xi_-)^{4}}\left ( \frac{\partial\xi_+}{\partial k_m}+\frac{\partial\xi_-}{\partial k_m} \right )
\right.\nonumber\\
+\frac{1}{(\xi_+-\xi_-)^3}\left.\left (\frac{\partial n(\xi_+)}{\partial \xi_+}\frac{\partial\xi_+}{\partial{k_m}}+ 
\frac{\partial n(\xi_-)}{\partial \xi_-}\frac{\partial\xi_-}{\partial{k_m}}   \right )\right ].~~
 \end{eqnarray}
Both integrals are determined by the integration over momentum space between the Fermi surfaces of two bands split  by the spin-orbital coupling. The phase transition to the superconducting state changes the Fermi distribution of the electrons over energies only in the narrow vicinities of the corresponding Fermi surfaces of the order of $\Delta$. Hence, the integration in eqs. (46) and (47)
leads to the result only slightly different from that is in the normal state. Even at zero temperature the relative magnitude of corrections do not exceed $\sim\Delta^2/\varepsilon_F^2$. So, 
the gyrotropy coefficient $\lambda$ determined by the integral $I_{lm}^3$ is practically keeps its normal state value given by eqn. (28) 
\begin{equation}
\lambda = i\frac{e^2\omega}{12\pi^2\gamma_0k_F}
 \left ( 1+{\cal O}({\Delta^2/\varepsilon_F^2})\right ).    
\end{equation}

The coefficient $\nu$ acquires nonzero value.  However,  it is much smaller than that was found in the paper Ref. 15.
The estimation made at $T=0$ yields
\begin{equation}
 \nu=\frac{e^2\gamma_0k_F}{\hbar^2c}{\cal O}(\Delta^2/\varepsilon_F^2)
\end{equation}

To find a relationship of the gyrotropy conductivity with observable optical properties one  has to develop 
electrodynamic theory of noncentrosymmetric metals. 

\section{Optical properties of noncentrosymmetric metal}

\subsection{Dispersion law}

To derive the light dispersion law we start from the Maxwell equations 
\begin{eqnarray}
\text{rot}{\bf B}=\frac{4\pi}{c}{\bf j}~,\\
\text{rot}{\bf E}=-\frac{1}{c}\frac{\partial {\bf B}}{\partial t}
\label{A2}
\end{eqnarray}
supplied by the density of current expression
\begin{equation}
{\bf j}=\frac{\epsilon}{4\pi}\frac{\partial {\bf E}}{\partial t}+\sigma{\bf E}+\lambda \text{rot}{\bf E}.
\end{equation}
The first term here corresponds to the dispersionless part of the displacement current.
The second one is the conductivity current written at infrared frequency region $\omega>v_F/\delta,~ \omega\tau>1$, where the current is locally related with an electric field, $\delta$ is the skin penetration depth. The last one is the gyrotropy current
\begin{equation}
{\bf j}_g=\lambda \text{rot} {\bf E}.
\label{j_g}
\end{equation}
As before we discuss the metal with the cubic point symmetry.

Eliminating the magnetic induction, we obtain
\begin{equation}
\nabla^2{\bf E}=\frac{\epsilon}{c^2}\frac{\partial^2 {\bf E}}{\partial t^2}+\frac{4\pi\sigma}{c^2}\frac{\partial {\bf E}}{\partial t}+\frac{4\pi\lambda}{c^2}\frac{\partial \text{rot}{\bf E}}{\partial t}.
\end{equation}
Taking solution for the circularly polarized wave
\begin{equation}
{\bf E}=(\hat x\pm i\hat y)E_0e^{i({\bf k}{\bf r}-\omega t)}
\end{equation}
we come to the dispersion relation
\begin{equation}
\label{k}
k^2=\frac{\epsilon\omega^2}{c^2}+\frac{4\pi i\sigma\omega}{c^2}\pm \frac{4\pi i\lambda\omega k}{c^2}.
\end{equation}
It is worth to be noted that for a media with time reversal breaking one has to substitute here $\sigma\to
\sigma_{\pm}=\sigma_{xx}\pm i\sigma_{xy}$, where $\sigma_{xy}$ is the Hall conductivity.

In neglect the gyrotropy term the complex index of refraction 
$$
 N=\frac{ck}{\omega}=n+i\kappa
 $$
is expressed through the diagonal part of complex conductivity $\sigma=\sigma'+i\sigma''$
by means of the usual relations 
$$
n^2-\kappa^2=\epsilon-\frac{4\pi\sigma''}{\omega},~~~~2n\kappa=\frac{4\pi\sigma'}{\omega}.$$
The gyrotropy term leads to  
 the difference in the refraction indices of clock wise and counter clock 
wise polarized light.
In the first order in respect to $\lambda=\lambda^{\prime}+i\lambda^{\prime\prime}$ the refraction index is
\begin{equation}
\label{N}
N^{\pm}=n+i\kappa\pm\frac{2\pi i \lambda}{c}.
\end{equation}
 Hence, the differences in the real  and imaginary parts of the refraction indices of circularly polarized lights  with the opposite 
polarization are
\begin{equation}
\Delta n=n_+-n_-=-\frac{4\pi\lambda''}{c},
\end{equation} 
\begin{equation}
\label{B}
\Delta\kappa=\kappa_+-\kappa_-=\frac{4\pi\lambda'}{c}.
\end{equation} 

In the superconducting state the gyrotropy current (\ref{j_g}) has more general form given by eqn. (\ref{g}). Hence, we should use the more general formula for the current
\begin{equation}
\label{C}
{\bf j}=\frac{\epsilon}{4\pi}\frac{\partial {\bf E}}{\partial t}+\sigma{\bf E}+\lambda \text{rot}{\bf E}+\nu {\bf B}.
\end{equation}
Then repeating all the calculations we come to the same results (\ref{k})-(\ref{B}) modified by  the substitution
\begin{equation}
\lambda~ \to~\Lambda=\lambda-\frac{ic\nu}{\omega}.
\end{equation}
We remind that the superconducting state current density given by eqn. (\ref{C}) is worth to use at  the high frequencies 
$\omega>v_F/\delta$ where the inequality $\hbar\omega>>\Delta$ is certainly valid. Here,  $\Delta$ is superconducting energy gap. In the low frequency limit $\hbar\omega<\Delta$ one should also  take into account the London density of current ${\bf j}_L=-(c/4\pi\delta_L^2)({\bf A}-\hbar c\nabla\varphi/2e)$. The interplay between the London current ${\bf j}_L$ and the Drude current ${\bf j}_D=\sigma(\omega){\bf E}$ is discussed in the textbook. \cite{FizKin}

\subsection{Magnetic moment}

The magnetization in gyrotropic media is 
\begin{equation}
{\bf M}=\frac{1}{2c}\lambda{\bf E},
\label{M_n}
\end{equation}
such that rotation of the magnetization is equal to one half of gyrotropy part of the current density
\begin{equation}
\frac{1}{2}~{\bf j}_g=c~\text{rot}{\bf M}.
\end{equation}
The relationship between the density of gyrotropy current and the magnetization 
is a general property of noncentrosymmetric materials (see also \cite{Levitov}). Both of them can be obtained
from the gyrotropy term in action 
$$
-\frac{1}{2c}\int dtd^3{\bf r}(\lambda{\bf E}){\bf B}.
$$
By variation of action in respect of $-{\bf B}$ and $-{\bf A}/c$, taking into account that $\lambda$ is an odd function of derivative $\partial/\partial t$ and making use the  definitions ${\bf E}=-(1/c) \partial{\bf A}/\partial t$ and ${\bf B}=\text{rot}
{\bf A}$,  we come to ${\bf M}$  given by eqn.(\ref{M_n}) and ${\bf j}_g$  given by eqn.(\ref{j_g})
 correspondingly.

All these considerations are valid both for the normal and as well for the superconducting state. However, in the latter case the gyrotropy action 
\begin{equation}
S_g=-\frac{1}{2c}\int dtd^3{\bf r}\left\{(\lambda{\bf E}){\bf B}
+\nu
\left [{\bf A}-\frac{\hbar c}{2e}\nabla \varphi\right ]{\bf B}\right\}.
\end{equation}
contains one extra term which is absent in the normal state due to the gauge invariance.
The corresponding expressions for the magnetic moment and gyrotropy current are
\begin{equation}
\label{M}
{\bf M}=\frac{1}{2c}\lambda{\bf E}+\frac{1}{2c}\nu\left [{\bf A}-\frac{\hbar c}{2e}\nabla \varphi\right ],
\label{M_s}
\end{equation}
\begin{equation}
{\bf j}_g=\lambda~\text{rot}{\bf E}+\nu{\bf B}.
\end{equation}

\subsection{Boundary conditions}

To consider the problem of light reflection normally incident to  the flat surface of noncentrosymmetric metal we need to find the relations between the wave amplitude propagating inside $(z>0)$ the material 
\begin{equation}
\label{in}
{\bf E}^{in}={\bf E}_0e^{i\omega(Nz/c-t)}
\end{equation}
and 
the amplitudes of incident and reflected waves outside it
\begin{equation}
\label{out}
{\bf E}^{out}={\bf E}_1e^{i\omega(z/c-t)}+{\bf E}_2e^{-i\omega(z/c+t)}.
\end{equation}
We have
$${\bf E}^{in}_{z=o}={\bf E}^{out}_{z=0}$$
that is
\begin{equation}
\label{E}
{\bf E}_0={\bf E}_1+{\bf E}_2.
\end{equation}
At the same time from the difference of the Maxwell equations (\ref{A2}) inside and outside of material 
we obtain
\begin{equation}
\label{rot}
(\text{rot}{\bf E}^{in}-\text{rot}{\bf E}^{out})_{z=0}=-\frac{1}{c}\frac{\partial }{\partial t}({\bf B}^{in}-{\bf B}^{out})_{z=0}.
\end{equation}
The difference of the magnetic inductions at the boundary is given by the jump of magnetzation
\begin{equation}
\label{DB}
({\bf B}^{in}-{\bf B}^{out})_{z=0}=4\pi{\bf M}_{z=0}.
\label{bc}
\end{equation}

In the stationary magnetic field parallel to the surface of the metal ${\bf H}^{out}=H_x\hat x$ this equation
yields
\begin{equation}
B^{int}_x(z=0)=H_x,~~~~B^{int}_y(z=0)=\frac{2\pi\nu}{c}A^{int}_y(z=0).
\label{bc}
\end{equation}
In the normal state where $\nu=0$ the boundary conditions add nothing special to the centrosymmetric case. In the superconducting state the solution of the London equations supplied by these boundary conditions results in quite unusual helical field distribution found in 
the paper. \cite{Levitov2}

For the light incident to the metallic surface using (\ref{rot}), (\ref{DB}), and (\ref{M}) we obtain
\begin{equation}
(\text{rot}{\bf E}^{in}-\text{rot}{\bf E}^{out})_{z=0}=-\frac{4\pi}{2c^2}\left (\lambda\frac{\partial }{\partial t}
-c\nu\right ){\bf E}^{in}_{z=0}
\end{equation}
Substituting here eqns. (\ref{in}), (\ref{out}) we come to
\begin{equation}
\hat z\times(N{\bf E}_0-{\bf E}_1+{\bf E}_2)=\frac{2\pi}{c}\Lambda{\bf E}_0
\end{equation}
For the combinations $E^\pm=E_x\pm iE_y$ of the electric field component  this relation can be rewritten as
\begin{equation}
\label{rotE}
\left (N^{\pm}\pm\frac{2\pi i\Lambda}{c}\right )E_0^{\pm}-E_1^{\pm}+E_2^{\pm}=0
\end{equation}

\subsection{Reflection coefficient and the Kerr effect} 

The equations (\ref{E}) and (\ref{rotE}) allow  express the amplitudes of reflected wave through the amplitude of the incident wave. We have for reflection coefficient
\begin{equation}
\label{R}
R^{\pm}=\frac{E_{2}^\pm}{E_{1}^\pm}=\frac{1-N^{\pm}\mp\frac{2\pi i\Lambda}{c}}
{1+N^{\pm}\pm\frac{2\pi i\Lambda}{c}},
\end{equation}
where the refraction index is
\begin{equation}
N_{\pm}=n+i\kappa\pm\frac{2\pi i \Lambda}{c}.
\end{equation}
Now one can rewrite eqn. (\ref{R}) in more habitual form 
\begin{equation}
R^{\pm}=\frac{1-\tilde N^{\pm}}
{1+\tilde N^{\pm}},
\end{equation}
where an effective refraction index is
\begin{equation}
\tilde N_{\pm}=n+i\kappa\pm\frac{4\pi i \Lambda}{c},
\end{equation}
and the effective differences in the real  and imaginary parts of the refraction indices of circularly polarized lights  with the opposite 
polarization are
\begin{equation}
\Delta \tilde n=\tilde n_+-\tilde n_-=-\frac{8\pi\Lambda''}{c},
\label{e44}
\end{equation} 
\begin{equation}
\Delta\tilde\kappa=\tilde\kappa_+-\tilde\kappa_-=\frac{8\pi\Lambda'}{c}.
\label{e45}
\end{equation}

Making use these definitions we can apply the standard procedure \cite{Ben} to calculate the Kerr rotation for linearly polarized light normally incident from vacuum to the flat boundary of a medium. 
The light is reflected as elliptically polarized with the major axis rotated relative to the incident polarization by an amount 
\begin{equation}
\theta=\frac{(1-n^2+\kappa^2)\Delta\tilde\kappa+2n\kappa\Delta \tilde n}{(1-n^2+\kappa^2)^2+(2n\kappa)^2}.
\label{e33}
\end{equation}  

\section{The Kerr rotation}

To find the Kerr rotation in the normal state let us substitute
 the eqn. (\ref{lambda})   in eqns. (\ref{e44}), (\ref{e45}). 
We find  $\Delta\tilde\kappa=0$ and  $\Delta\tilde n$
expresses through ratio of the light frequency to the band splitting $2\gamma_0 k_F$ as
\begin{equation}
\Delta \tilde n=-\frac{2\alpha}{3\pi}\frac{\hbar\omega}{\gamma_0 k_F}.
\label{e6}
\end{equation} 
Here, $\alpha=e^2/\hbar c$ is the fine structure constant.

We limit ourselves by the frequencies not
exceeding the band splitting $\gamma_0 k_F$. Although the latter is not known for many noncentrosymmetric materials, one can expect it is about
 hundred Kelvin or in the frequency units   $\sim ~10^{13} rad/sec$.\cite{SZB04}  As  an example  we  consider the 
situation when the frequency of light is of the order of this value and larger than the quasiparticles scattering rate (clean limit): $1<<\omega\tau<\omega_p\tau$, where $\omega_p=\sqrt{4\pi ne^2/m^*}$
is the plasma frequency.  In this frequency region the real and imaginary part of conductivity 
are $\sigma'\approx\omega_p^2/4\pi \omega^2\tau$ and $\sigma''\approx\omega_p^2/4\pi \omega$.
Then,  one can find $2n\kappa\approx\omega_p^2/\omega^3\tau$ and $\kappa^2-n^2\approx
\omega_p^2/\omega^2$. Thus, for the Kerr angle we obtain
\begin {equation}
\theta\approx
-\frac{2\alpha}{3\pi}\frac{\hbar\omega^2}{\gamma_0 k_F\omega_p^2\tau}.
\label{e3}
\end{equation} 
So, the Kerr angle in noncentrosymmetric metals can have measurable magnitude, in particular if we compare it
 with
the Kerr angle of the order of $6\times 10^{-8} ~rad$
measured  in the superconducting $Sr_2RuO_4$ by the Stanford group.\cite{Xia}

For $\Delta\tilde n$ in the superconducting state we obtain
\begin{equation}
\Delta \tilde n_s=-\frac{2\alpha}{3\pi}
  \left[
    \frac{\hbar\omega}{\gamma_0 k_F}
    \left( 1+{\cal O}(\Delta^2/\varepsilon_F^2)\right)
   -\frac{\gamma_0k_F}{\hbar \omega}
    {\cal O}(\Delta^2/\varepsilon_F^2)  \right].  
           \end{equation}
Finally, for the Kerr angle in the same frequency interval as for the normal state we have
\begin{equation}
 \theta_s=-\frac{2\alpha\omega}{3\pi \omega_p^2\tau}
  \left[
    \frac{\hbar\omega}{\gamma_0 k_F}
    \left( 1+ {\cal O}(\Delta^2/\varepsilon_F^2)\right)
   -\frac{\gamma_0k_F}{\hbar \omega}
    {\cal O}(\Delta^2/\varepsilon_F^2)\right].               
\end{equation}

\section{Conclusion}

We have presented here the derivation of the current response 
to the electromagnetic field with finite frequency and wave vector in noncentrosymmetric metal. 
The obtained general formula valid both in the normal and in the superconducting state was applied to the calculation of observable physical properties 
in the frequency interval smaller than the band splitting $\hbar\omega< \gamma_0k_F$.
The latter in its turn was supposed to be smaller than the Fermi energy $\gamma_0k_F<\varepsilon_F$.
The calculations was performed in the clean case $\omega\tau>1$, that is, in particular, important to neglect the vortex corrections.  We did not discuss the  anomalous skin effect assuming 
that the wave length does not exceed the skin penetration depth $\delta>v_F/\omega$.
In the normal state the current contains the gyrotropic part which is odd function of the wave vector and the frequency.
It presents a sort of displacement  current originating of band splitting in noncentrosymmetric metal.
In the superconducting state there is an additional part of the gyrotropy current proportional to magnetic field. The change of the gyrotropy conductivity
 in the superconducting state was found.
As an example the Kerr rotation for the polarized light reflected from the surface of noncentrosymmetric metal with cubic symmetry is calculated.

\acknowledgments

One of the authors (V. P. M.) is indebted to E. Kats  and L. Falkovsky for the numerous helpful discussions of natural optical activity and technical problems related to its calculation.

The financial support of  another author (Y. Y.) by the Global COE program (G10) from Japan Society for the Promotion of Science is gratefully acknowledged.


\begin{thebibliography}{99}
\bibitem{neglect} The estimation of neglected terms shows that they yield small corrections of the order of $\gamma_0k_F/\varepsilon_F$ to the main terms determining the gyrotropy current.
\end{thebibliography}
\end{document}


\title{Erratum: Optical activity of noncentrosymmetric metals [Phys. Rev.B {\bf 81}, 094525 (2010)],
Magnetostatic and optics of noncentrosymmetric metals [Phys.Rev.B {\bf 88}, 134514 (2013)]}

\author{V.P.Mineev}
\affiliation{Commissariat a l'Energie Atomique, INAC / SPSMS, 38054 Grenoble, France}


\pacs{ 78.20.Ek,74.20.Fg,74.25.N-,74.72.-h}

\date{\today}
\maketitle

The boundary condition for the combinations $E^\pm=E_x\pm iE_y$ of the electric field component given by Eq. (75) in the paper \cite{Mineev2010} should be
\begin{equation}
\label{rotE}
\left (N^{\pm}\mp\frac{2\pi i\Lambda}{c}\right )E_0^{\pm}-E_1^{\pm}+E_2^{\pm}=0.
\end{equation}
Sign in the bracket in this equation  is $\mp$ unlike to the sign $\pm$ in the paper \cite{Mineev2010}. The wrong sign has originated from an arithmetic error in the transition from Eq. (74) to Eq. (75) in\cite{Mineev2010}.
Here, $E_1$ and $E_2$ correspondingly are the complex amplitudes of electric component of em wave normally incident and reflected from  the surface of noncentrosymmetric media, $E_0$ is the amplitude of transmitted wave. $N^{\pm}=N\pm\frac{2\pi i\lambda}{c}$ is refractive index of gyrotropic media.
 The second boundary condition is given by Eq. (69)
\begin{equation}
E_0^\pm=E_1^\pm+E_2^\pm.
\end{equation}
The solution of these two equations yields the reflection coefficients for the clockwise and counter clockwise circular light polarization
\begin{equation}
\label{R}
R^{\pm}=\frac{E_{2}^\pm}{E_{1}^\pm}=\frac{1-N}
{1+N}.
\end{equation}
Thus, unlike to the statement  formulated in \cite{Mineev2010} $R^+$ is proved to be equal to $R^-$. Hence, the Kerr effect  for the rotation of polarization of light reflected from the naturally gyrotropic media found in papers \cite{Mineev2010,Mineev2013} is in fact absent.

In microscopic calculations of  Kerr rotation undertaken  in the papers \cite{Hosur, Orenstein,Pershiguba} were used the phenomenological expression for the Kerr angle obtained by A.R.Bungay, Yu.P.Swirko, and N.I. Zheludev\cite{Bungay}. These authors  have derived boundary conditions for electric field making use the textbook procedure and the expression for the gyrotropy current  given by Eq. (30) in paper \cite{Bungay} which in our notations looks as follows
\begin{equation}
{\bf j}_g=~\text{rot}{\lambda\bf E}.
\end{equation}
In fact the correct  gyrotropy current expression can be obtained 
from the gyrotropy term in action 
\begin{equation}
-\frac{1}{2c}\int dtd^3{\bf r}(\lambda{\bf E}){\bf B}.
\label{action}
\end{equation}
By variation of action in respect of $-{\bf A}/c$
and making use the  definitions ${\bf E}=-(1/c) \partial{\bf A}/\partial t$ and ${\bf B}=\text{rot}
{\bf A}$,  we come to 
\begin{equation}
{\bf j}_g=~\text{rot}{\lambda\bf E}-\frac{1}{2}\nabla\lambda\times{\bf E}.
\label{g}
\end{equation}
Here , unlike to \cite{Mineev2010} we have taken into account coordinate dependence of $\lambda$ term 
$$
\lambda(z)=\lambda\theta(z)
$$
corresponding to half a space $(z>0)$  filled by gyrotropic media.
The standard derivation making use the Maxwell equations and the gyrotropy current (\ref{g}) leads to the boundary conditions (\ref{rotE}) and, hence, to the Kerr effect absence. It is worth to be noted
that formula (\ref{rotE})  in the paper \cite{Mineev2010} has been derived by different way not using the formula (\ref{g}) for gyrotropy current but the expression for the gyrotropy magnetic moment
\begin{equation}
{\bf M}_g=\frac{1}{2c}\lambda{\bf E}
\label{M_n}
\end{equation}
also obtained from the gyrotropy action Eq. (\ref{action}). Thus, now both derivations of boundary conditions Eq.(1) yield the same result leading to the absence of the Kerr rotation  at   light reflection from  media with broken space inversion. On the contrary the Kerr effect arises at reflection from  media with broken time inversion. So, to explain the Kerr effect observation in high-$T_c$ cuprates (see numerous references and discussion in the paper \cite{Hosur}) one must to apply more efforts.

In conclusion, I express my gratitudes to   Weejee Cho who pointed out correct $\mp$ sign in the Eq.(1),
as well to A.D.Fried for the valuable discussions and to R.B.Laughlin for the deep remark.